\documentclass{article}

\usepackage{arxiv}

\usepackage[utf8]{inputenc} 
\usepackage[T1]{fontenc}    
\usepackage{hyperref}       
\usepackage{url}            
\usepackage{booktabs}       
\usepackage{amsfonts}       
\usepackage{nicefrac}       
\usepackage{microtype}      
\usepackage{lipsum}		
\usepackage{graphicx}
\usepackage[square,sort&compress,comma,numbers]{natbib}
\usepackage{doi}

\title{Designing Illumination Patterns for Single-Pixel Imaging Using Lattice Models}

\author{ \href{https://orcid.org/0000-0002-5705-5788}{\includegraphics[scale=0.06]{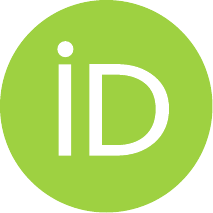}\hspace{1mm}Hamidreza Oliaei-Moghadam}\thanks{Email: \texttt{oliaeimoghaddam@gmail.com}} \\
	Department of Physics\\
	Shiraz University\\
	 71946-84795, Fars, Iran \\
}

\renewcommand{\shorttitle}{\textit{arXiv} Template}

\hypersetup{
pdftitle={Designing Illumination Patterns for Single-Pixel Imaging Using Lattice Models},
pdfsubject={q-bio.NC, q-bio.QM},
pdfauthor={David S.~Hippocampus, Elias D.~Striatum},
pdfkeywords={First keyword, Second keyword, More},
}

\begin{document}
\maketitle

\begin{abstract}
Single-pixel imaging leverages a single-pixel detector and structured illumination patterns to reconstruct images, offering a cost-effective solution for imaging across a wide range of wavelengths, such as x-ray and terahertz. However, the technique faces challenges in efficiency due to the need for numerous patterns to achieve high-quality image reconstruction. In this study, we explore the use of spin lattice models from statistical mechanics to design illumination patterns for single-pixel imaging. By employing models like Ising, Potts, XY, and Heisenberg, we generate structured patterns that are adaptable for binary, grayscale, and color imaging. This work creates a direct connection between lattice models and imaging applications, providing a systematic approach to pattern generation that can enhance single-pixel imaging efficiency.
\end{abstract}


\section{Introduction}
Single-pixel imaging offers a unique approach to capturing images by utilizing a single-pixel detector rather than an array detector, as is typical in modern cameras. Instead of directly capturing the scene, single-pixel cameras use a sequence of mask patterns to filter the scene and record the transmitted intensity for each pattern \cite{sun2019single,gibson2020single}. This technique enables the development of cost-effective imaging systems across various wavelengths of the electromagnetic spectrum, such as x-ray imaging \cite{klein2019x,he2020high}, terahertz imaging \cite{shrekenhamer2013terahertz}, and fluorescence microscopy\cite{field2016single}. A straightforward way to capture an image with a single-pixel detector is to measure each pixel sequentially. However, this method is inefficient in utilizing available illumination. A more efficient strategy involves using a sequence of spatially resolved patterns and recording the intensity measurements that correlate with these patterns and the scene. The resulting image is reconstructed using the following correlation formula \cite{gibson2020single,Simon2017,edgar2019principles}:

\begin{equation}
    T'(x,y)_{GI} = \frac{1}{N} \sum_{r=1}^{N} \left( B_{r} - \langle B \rangle \right) I_{r}(x,y),
\end{equation}

where \( T'(x,y) \) represents the image, \( B_{r} \) is the total intensity recorded by the single-pixel detector, and \( I_{r}(x,y) \) is the pattern function for the \( r \)-th measurement.

This correlation-based imaging can be implemented in two ways\cite{gibson2020single,qiu2020comprehensive,shapiro2008computational}:
\begin{itemize}
    \item \textbf{Structured Detection:} A light modulator in the image plane of a camera lens masks images of the scene, and the single-pixel detector measures the filtered intensities.
    \item \textbf{Structured Illumination:} The light modulator projects patterns onto the scene, and the single-pixel detector captures the backscattered intensities.
\end{itemize}

Although effective, this imaging method faces a major limitation: achieving high-quality images requires a large number of patterns, which can significantly slow down the process.

The speed of this imaging technique can be improved through\cite{duarte2008single,zhang2015single,zhang2017hadamard,bian2017experimental}:
\begin{itemize}
    \item Developing advanced hardware for faster pattern generation or light detection.
    \item Using more optimized patterns for image construction.
    \item Implementing improved reconstruction algorithms.
\end{itemize}

Statistical mechanics plays a crucial role in both pattern generation and reconstruction algorithms.
Although deep learning models, as data-driven approaches, have been extensively applied in this field, theory-driven models have received comparatively little attention. However, theory-driven models hold considerable potential for further optimization. In particular, spin lattice models (e.g., Ising, Potts, XY, Heisenberg) can be effectively employed to design optimal illumination patterns.
These patterns are well-suited for imaging because the correlation between adjacent spins in lattice models mimics the natural correlation between adjacent pixels in real images. In this paper, we propose methods for employing lattice models to create binary, grayscale, indexed color, and full-color patterns, enhancing the efficiency and quality of single-pixel imaging.

\section{A Brief Overview of Lattice Models}
In statistical mechanics, spin lattice models represent a distinct class of systems where spins are arranged on a lattice and interact with their neighboring spins. These models can often be viewed as idealized representations of magnets, providing a framework for describing numerous magnetic phenomena in materials. Classical spin systems, such as these, play a key role in understanding magnetism and phase transitions.
Here, we provide an overview of some of these models, focusing on their potential applications in generating patterns.
\subsection{Ising model}
The Ising model is a well-known statistical mechanics model used to describe ferromagnetism. In its two-dimensional form, it employs discrete variables
$ s_i = \pm 1 $, which are defined on a square lattice and represent the spin value at those sites. These spins can be in either the "up" ($ +1 $) or "down" ($ -1 $) state.
Assuming the presence of an external magnetic field denoted by $ H $ and considering the interaction of each spin with its four nearest neighbors, the energy of a specific configuration, represented as $ \{ s_1, s_2, \dots, s_N \} $, is defined using the Hamiltonian function $ E $ as follows\cite{Pathria2022,Herbut2007,cipra1987introduction,newell1953theory}:  

\begin{equation}
    E=-J\sum_{\langle i,j \rangle}s_is_j - H\sum_{i=1}^{N}s_i,
    \label{eg:energyIsing}
\end{equation}
where $ J>0 $ is the spin-spin coupling constant, and the notation $ \langle i,j \rangle $ indicates that the summation in the Hamiltonian accounts only for interactions between adjacent spins in the lattice.  
If the external field is absent ($ H = 0 $), the Hamiltonian simplifies to:  
\begin{equation}
    E=-J\sum_{\langle i,j \rangle}s_is_j.
    \label{eg:energyIsing0field}
\end{equation}
If the coupling constant in the Hamiltonian (\ref{eg:energyIsing0field}) is $ J < 0 $, the system is referred to as an antiferromagnet. In an antiferromagnetic system, neighboring spins tend to align in opposite directions.
The Hamiltonian (\ref{eg:energyIsing0field}) can be generalized to:  
\begin{equation}
    E=-\sum_{\langle i,j \rangle} J_{i,j} s_i s_j.
    \label{eg:Anderson}
\end{equation}
where $ J_{i,j} $ can take arbitrary positive or negative values for each pair of spin. This model is known as the Edwards-Anderson model.  
By adding a site-dependent magnetic field to the Edwards-Anderson model, the following Hamiltonian is obtained:  
\begin{equation}
    E=-\sum_{\langle i,j \rangle} J_{i,j} s_i s_j -\sum_i h_{i} s_i.
    \label{eg:GeneralIsing}
\end{equation}
Another way to generalize Hamiltonian (\ref{eg:energyIsing0field}) is when we allow each spin to have more than 2 values.
\subsection{Potts model}
The Potts model is a generalization of the Ising model that allows each spin to have multiple values, typically more than two. The standard Potts Hamiltonian is given by \cite{wu1982potts,wu1984potts,mcdonald2012potts}:

\begin{equation}
    E = -J_p \sum_{\langle i,j \rangle} \delta(s_i, s_j),
    \label{eq:potts_Hamiltonian}
\end{equation}

\begin{figure}[!ht]
	\centering
	\includegraphics[width=.3\textwidth]{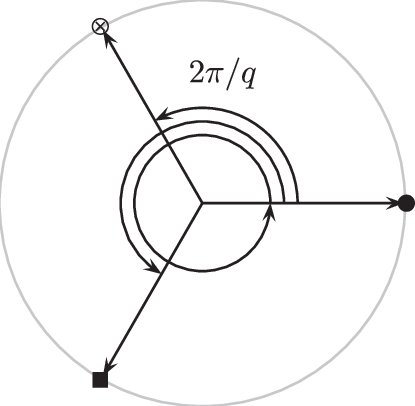}
\caption{
    Angles in a Potts model with $q$ allowed states. 
    In this example, $q=3$ is chosen, and the angles are given by $0, 2\pi/3, 4\pi/3$
\cite{wipf2007generalized}
.
}
	\label{fig:PottsVector}
\end{figure}

Where $\delta(s_i, s_j)$ is the Kronecker delta function, which is equal to 1 if $s_i = s_j$ and 0 otherwise. In this model, the variable $s_i$ represents the spin configuration at a lattice site. The allowed values for $s_i$ are restricted to nonnegative integers from $0$ to $q-1$, where $q$ denotes the number of available states. For example, in the case of $q=3$, the spin variable $s_i$ can take values from the set $\{0, 1, 2\}$.  
Another version of the Potts model is called the vector Potts model or clock model, which consists of spins arranged on a lattice, typically a two-dimensional rectangular lattice, though it can be generalized to different dimensions and lattice structures.
In the vector Potts model, the spin values are uniformly distributed around a circle at angles given by $\theta_s = 2\pi s / q$ for $s = 0,1,\dots,q-1$ (Figure \ref{fig:PottsVector}).
The interaction Hamiltonian in this model is given by:  
\begin{equation}
E= -J_c \sum_{\left \langle i,j \right \rangle} \cos(\theta_{s_i} - \theta_{s_j}).
\end{equation}
It is worth noting that in the limit of $ q \rightarrow \infty $, this model becomes the $ XY $ model, which will be discussed next.

\subsection{XY model}
The XY model is a two-dimensional spin model that is an extension of the vector Potts model. Unlike the Potts model, where spins can only take discrete values, in the XY model, spins can be oriented in any direction within a plane. This feature makes the XY model a more accurate representation of certain physical systems, such as thin magnetic layers.
In the XY model, the system's energy is defined by \cite{barouch1970statistical,kosterlitz1974critical,gupta1988phase}:
\[
E = -J \sum_{\langle i,j \rangle} \cos(\theta_i - \theta_j).
\]
In addition, \( \theta_i \) and \( \theta_j \) represent the angles of the spins at the sites \(i\) and \(j\), respectively. The factor \( \cos(\theta_i - \theta_j) \) represents the energy contribution of each pair of spins, where the energy decreases as the spins align. If the spin orientations in the XY model lattice are not confined to a plane, the system can instead be described by the Heisenberg model.
\subsection{Classical Heisenberg model}
The classical Heisenberg model, is described by the Hamiltonian function presented in Equation (\ref{eqn:Heisenberg}). This model, which belongs to the class of $n$-vector models, is commonly used in statistical mechanics to explain phenomena such as ferromagnetism. It was developed by Werner Heisenberg and plays a significant role in understanding spin interactions in many-body systems \cite{joyce1967classical,stanley1966possibility}:
\begin{equation}
	H = -J  \sum_{\langle i,j \rangle}\mathbf{S}_i \cdot \mathbf{S}_j,
	\label{eqn:Heisenberg}
\end{equation}
where
$\mathbf{S}_i = (s_x,s_y,s_z)_i$ 
represents the spin vector at the position $ i $ in the lattice site. The symbol $ \cdot $ denotes the dot product between two spins and $ J $ is the interaction coefficient between them.
\section{Creating Illumination patterns using lattice models}
Time-efficient single-pixel imageing strongly depends on the generation of effective illumination patterns. In this section, we explain how different models contribute to the creation of diverse illumination patterns for imaging purposes.
\subsection{Binary patterns}
\begin{figure}[!ht]
 	\centering
 	\includegraphics[width=\textwidth]{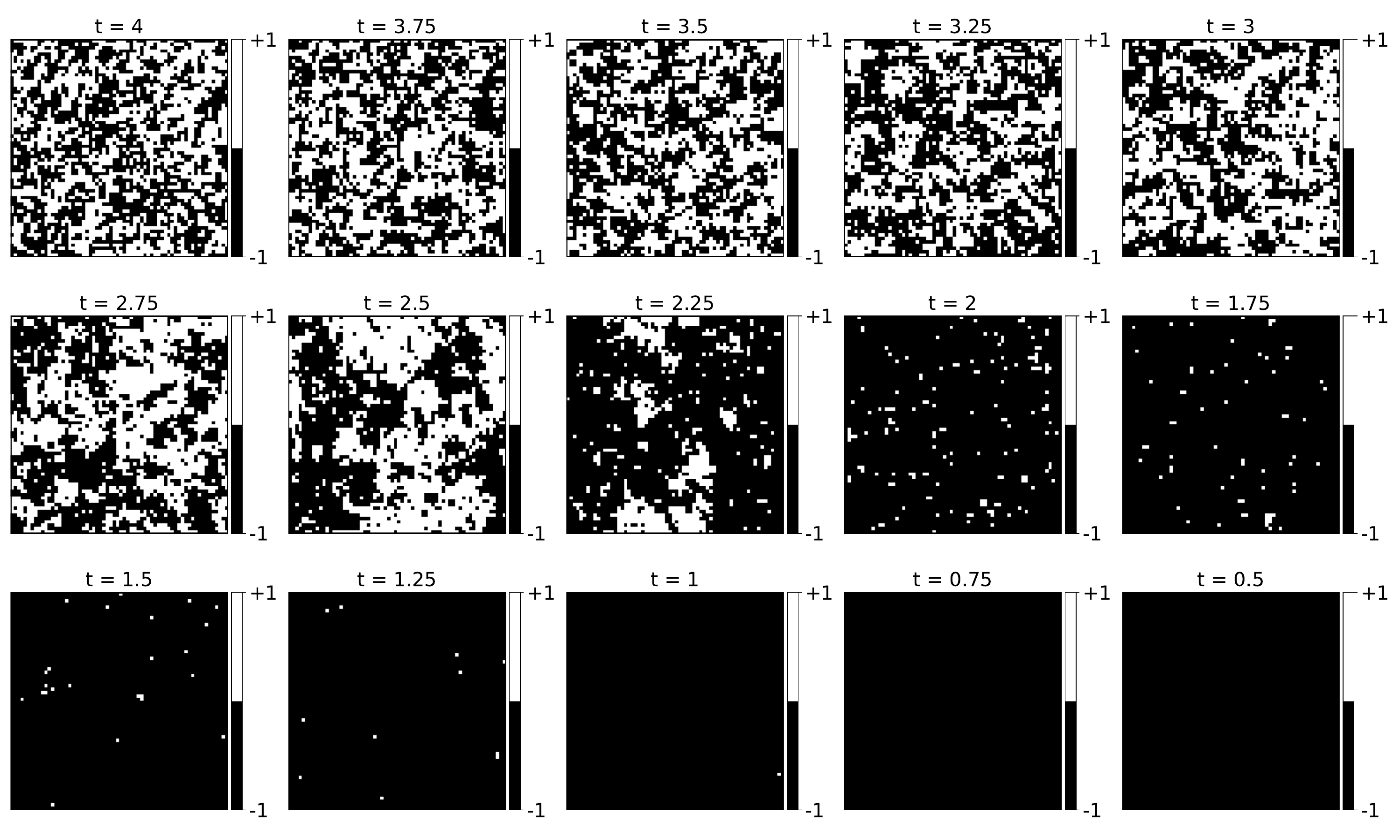}
 	\caption{The effect of temperature decrease on the spin configuration in the Ising model. Here, $t = (k_B T)/J$}
 	\label{fig:IsingPatterns}
 \end{figure}

Binary patterns are patterns in which each pixel has only two states: on or off, and the illumination intensity of each pixel remains constant. Given the binary nature of each pixel, the Ising model is a suitable candidate to generate such patterns. In the Ising model, the spin values can take two possible values: $-1$ and $+1$. 
When using the Ising model Hamiltonian to generate patterns, the characteristics of the generated patterns depend on the couplings between spins and the external fields. These couplings and fields can be constant values, functions of position on the lattice, or functions of temperature. Therefore, the desired pattern can be generated first at a specific temperature and field, and then the value $-1$ can be considered equivalent to a pixel being off, while the value $+1$ can be considered equivalent to a pixel being on. For example, assuming the external field is zero, the changes in a specific spin system are shown in Figure \ref{fig:IsingPatterns} as the system's temperature decreases from high to low.
As observed, as the temperature decreases, the size of the spots in the spin arrangement increases, while the total magnetization of the system remains close to zero. However, when the temperature drops below a certain threshold, known as the critical temperature, the system undergoes a phase transition to an ordered phase, where spins tend to align in the same direction (in this case, the spins are aligned in the -1 direction).

\begin{figure}[!ht]
	\centering
	\includegraphics[width=\textwidth]{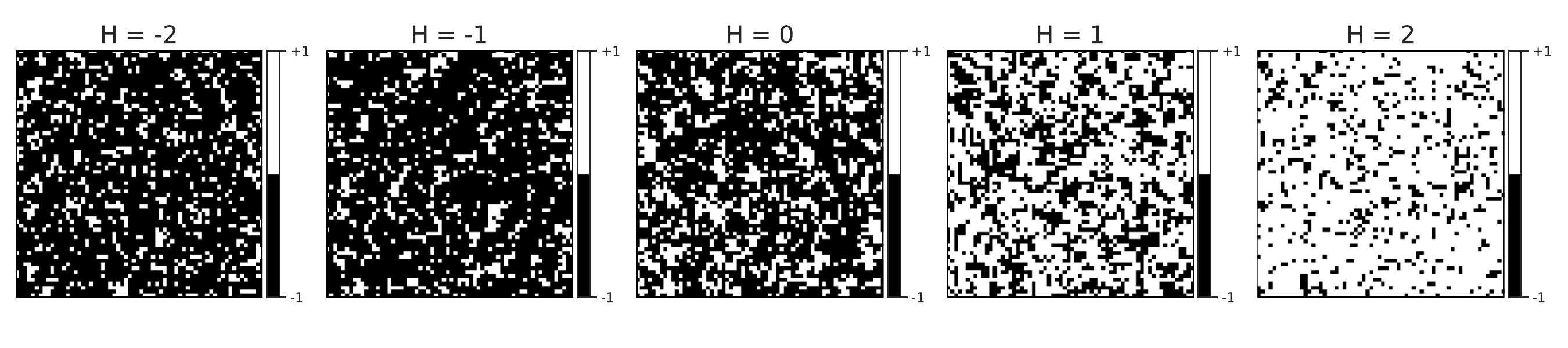}
\caption{Spin configuration in the Ising model at temperature $ t=4 $ with different magnetic field values applied to the spin system. Applying a negative (positive) field causes a larger proportion of spins to have the value $ -1 (+1) $, resulting in a non-zero magnetization ($ m \neq 0 $).}
\label{fig:Ising_fielded}
\end{figure}

In another example, we applied a non-zero external field at a constant temperature, which can be either positive or negative, as shown in Figure \ref{fig:Ising_fielded}.
It can be observed that by controlling the external field and temperature, the size and density of spots in a pattern can be controlled. Additionally, using the information obtained during the imaging process, different strategies can be employed to generate patterns suitable for single-pixel imaging. One of these strategies is discussed in \cite{oliaei2023patterns}.

\subsection{Grayscale Patterns and indexed color patterns} 
\begin{figure}[!ht]
	\centering
	\includegraphics[width=\textwidth]{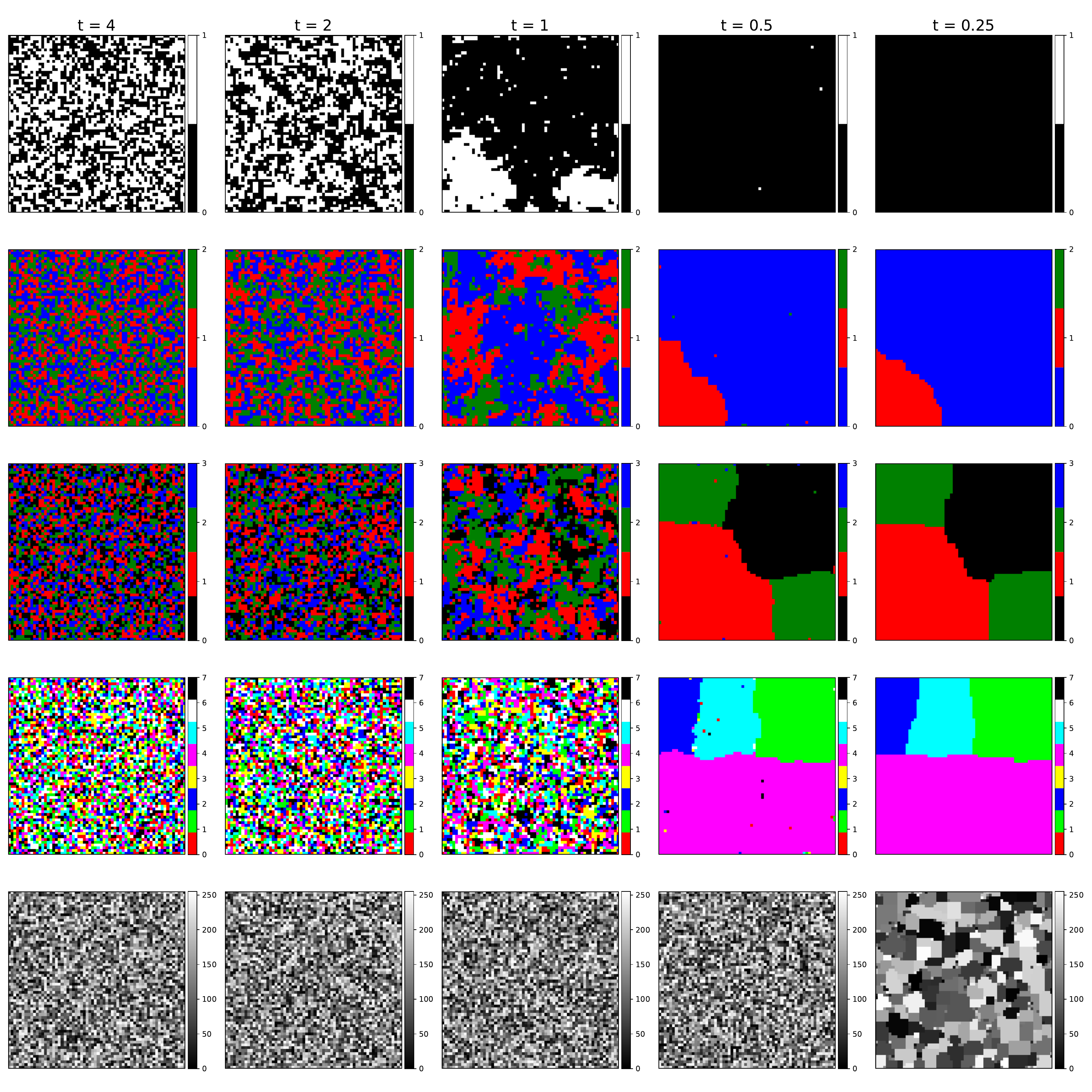}
	\caption{The effect of temperature reduction on the spin configuration in the standard Potts model. Here, $t = (k_B T)/J$.
Each row in the figure corresponds to different systems with varying numbers of allowed states (q). For example, the first row shows the temperature evolution of a 2-state vector Potts system, while the last row illustrates the temperature evolution of a 256-state vector Potts system.}
	\label{fig:PottsPatterns}
\end{figure}

If the spin values in the Potts model are assigned to specific colors, the resulting patterns can be used as indexed-color patterns.
Full-spectrum grayscale illumination patterns can be generated using the Potts model with 256 states, where each state corresponds to a distinct light intensity.  These patterns, encompassing all shades of gray, are essential for representing grayscale images in imaging applications. Figure (\ref{fig:PottsPatterns}) displays a selection of patterns generated using the standard two-dimensional Potts model for q = 2, 3, 4, 8, and 256 at various temperatures.
\par
\begin{figure}[!ht]
	\centering
	\includegraphics[width=\textwidth]{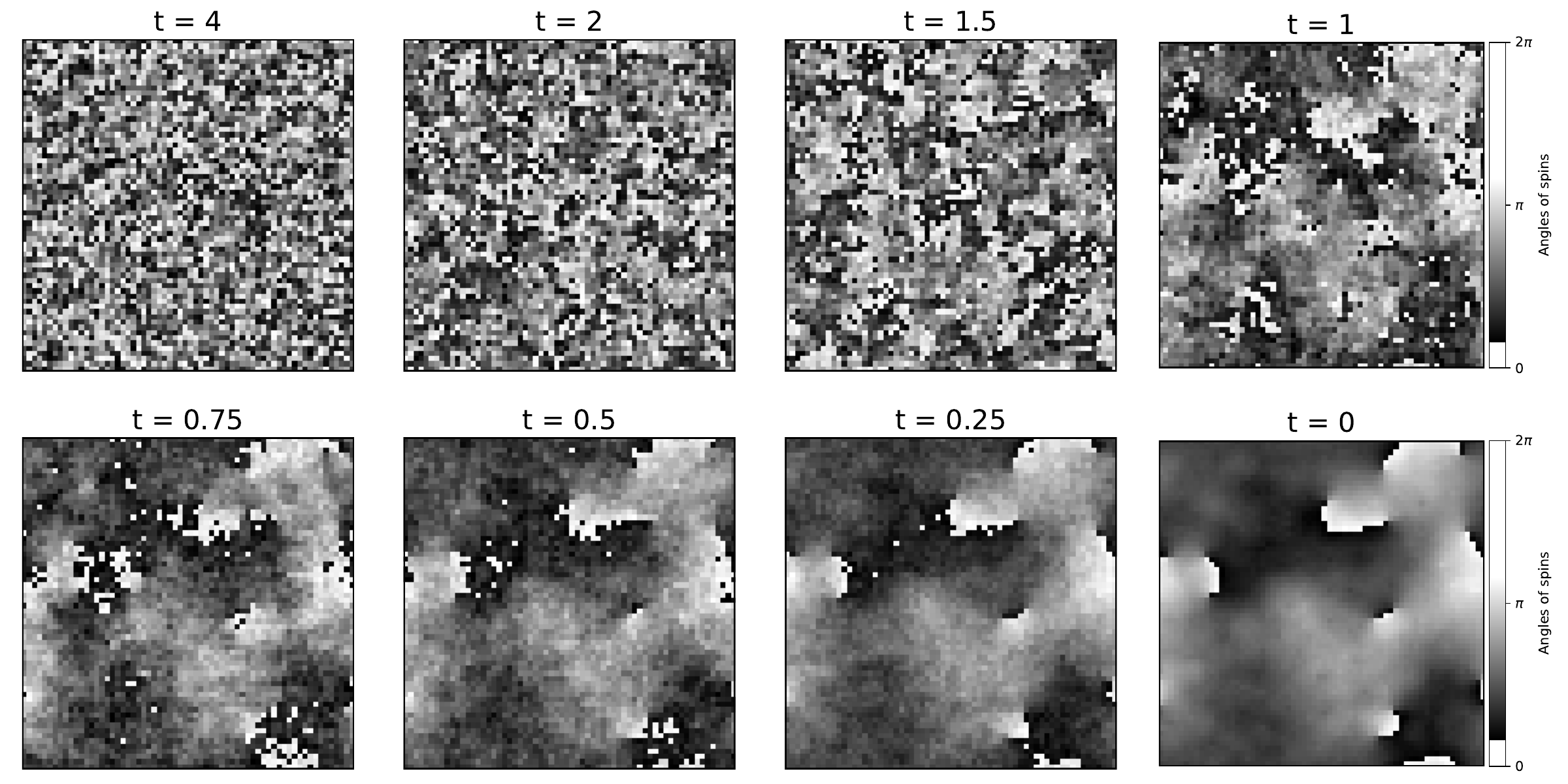}
\caption{The effect of temperature reduction on spin configuration in the XY model. Here, $t = (k_B T)/J$.}
	\label{fig:xy_gray}
\end{figure}

Alternatively, the two-dimensional XY model can generate grayscale patterns by mapping rotation angles at each pixel to grayscale intensities. This approach provides a flexible and distinct method for grayscale pattern design. Figure \ref{fig:xy_gray} shows an example of an XY model pattern and its evolution with decreasing temperature.
\subsection{Color patterns}
\begin{figure}[!ht]
	\centering
	\includegraphics[width=\textwidth]{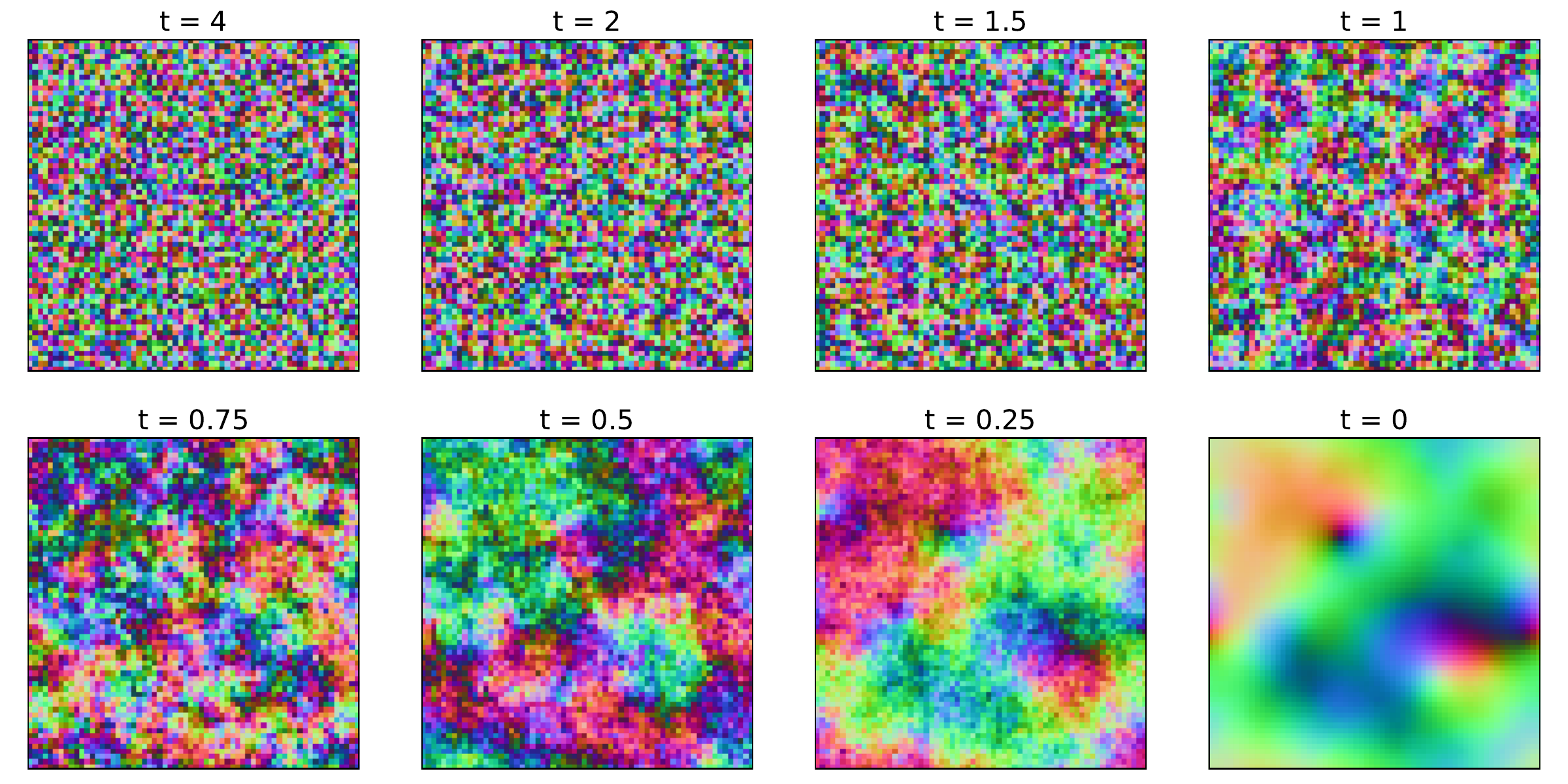}
\caption{The effect of temperature reduction on the arrangement of spins in the Heisenberg model. The spins are represented by components $S_x$, $S_y$, and $S_z$, visualized using an RGB color scheme. The color at each lattice site corresponds to the spin orientation in the $x$, $y$, and $z$ directions. As the temperature decreases ($t = \frac{k_B T}{J}$), the evolution of spin patterns can be observed from left to right and top to bottom.}
	\label{fig:HeisenbergPattern}
\end{figure}

To create the full spectrum of possible color patterns, the Heisenberg model can be employed.  Specifically, the spin values computed along the x, y, and z axes are scaled and then mapped directly to the corresponding red, green, and blue channels of the RGB color space. This process enables the generation of complex illumination patterns with nuanced variations in both color and intensity. An example of a Heisenberg model pattern and its changes as the temperature is reduced is shown in Figure \ref{fig:HeisenbergPattern}.
\section{Conclusion}
In this paper, we propose a new framework for creating illumination patterns in single-pixel imaging using spin lattice models. Models such as Ising, Potts, XY, and Heisenberg systems are used to generate patterns that can be applied to binary, grayscale, and color imaging. This novel approach adds a layer of structure and methodology to the design of illumination patterns, offering a fresh perspective on improving the efficiency and versatility of single-pixel imaging techniques.
\clearpage
\bibliographystyle{unsrtnat}


\begin{thebibliography}{1}
\bibitem{gibson2020single}
Gibson, G. M., Johnson, S. D., \& Padgett, M. J. (2020). Single-pixel imaging 12 years on: a review. Optics express, 28(19), 28190-28208.
\bibitem{edgar2019principles}
Edgar, M. P., Gibson, G. M., \& Padgett, M. J. (2019). Principles and prospects for single-pixel imaging. Nature photonics, 13(1), 13-20.

\bibitem{sun2019single}
Sun, M. J., \& Zhang, J. M. (2019). Single-pixel imaging and its application in three-dimensional reconstruction: a brief review. Sensors, 19(3), 732.


\bibitem{klein2019x}
Klein, Y., Schori, A., Dolbnya, I. P., Sawhney, K., \& Shwartz, S. (2019). X-ray computational ghost imaging with single-pixel detector. Optics express, 27(3), 3284-3293.

\bibitem{he2020high}
He, Y. H., Zhang, A. X., Li, M. F., Huang, Y. Y., Quan, B. G., Li, D. Z., ... \& Chen, L. M. (2020). High-resolution sub-sampling incoherent x-ray imaging with a single-pixel detector. APL Photonics, 5(5).

\bibitem{shrekenhamer2013terahertz}
Shrekenhamer, D., Watts, C. M., \& Padilla, W. J. (2013). Terahertz single pixel imaging with an optically controlled dynamic spatial light modulator. Optics express, 21(10), 12507-12518.

\bibitem{field2016single}
Field, J. J., Winters, D. G., \& Bartels, R. A. (2016). Single-pixel fluorescent imaging with temporally labeled illumination patterns. Optica, 3(9), 971-974.


\bibitem{Simon2017}
Simon, D. S., Jaeger, G., Sergienko, A. V., Simon, D. S., Jaeger, G., \& Sergienko, A. V. (2017). Quantum metrology (pp. 91-112). Springer International Publishing.

\bibitem{qiu2020comprehensive}
Qiu, Z., Zhang, Z., \& Zhong, J. (2020). Comprehensive comparison of single-pixel imaging methods. Optics and Lasers in Engineering, 134, 106301.

\bibitem{shapiro2008computational}
Shapiro, J. H. (2008). Computational ghost imaging. Physical Review A—Atomic, Molecular, and Optical Physics, 78(6), 061802.

\bibitem{duarte2008single}
Duarte, M. F., Davenport, M. A., Takhar, D., Laska, J. N., Sun, T., Kelly, K. F., \& Baraniuk, R. G. (2008). Single-pixel imaging via compressive sampling. IEEE signal processing magazine, 25(2), 83-91.
\bibitem{zhang2015single}
Zhang, Z., Ma, X., \& Zhong, J. (2015). Single-pixel imaging by means of Fourier spectrum acquisition. Nature communications, 6(1), 6225.
\bibitem{zhang2017hadamard}
Zhang, Z., Wang, X., Zheng, G., \& Zhong, J. (2017). Hadamard single-pixel imaging versus Fourier single-pixel imaging. Optics Express, 25(16), 19619-19639.

\bibitem{bian2017experimental}
Bian, L., Suo, J., Dai, Q., \& Chen, F. (2017). Experimental comparison of single-pixel imaging algorithms. Journal of the Optical Society of America A, 35(1), 78-87.

\bibitem{wu1982potts}
Wu, F. Y. (1982). The potts model. Reviews of modern physics, 54(1), 235.
\bibitem{mcdonald2012potts}
McDonald, L. M., \& Moffatt, I. (2012). On the Potts model partition function in an external field. Journal of Statistical Physics, 146, 1288-1302.
\bibitem{wipf2007generalized}
Wipf, A., Heinzl, T., Kaestner, T., \& Wozar, C. (2007). Generalized Potts-models and their relevance for gauge theories. SIGMA. Symmetry, Integrability and Geometry: Methods and Applications, 3, 006.
\bibitem{barouch1970statistical}
Barouch, E., McCoy, B. M., \& Dresden, M. (1970). Statistical mechanics of the XY model. I. Physical Review A, 2(3), 1075.
\bibitem{wu1984potts}
Wu, F. Y. (1984). Potts model of magnetism. Journal of Applied Physics, 55(6), 2421-2425.

\bibitem{gupta1988phase}
Gupta, R., DeLapp, J., Batrouni, G. G., Fox, G. C., Baillie, C. F., \& Apostolakis, J. (1988). Phase transition in the 2 D XY model. Physical review letters, 61(17), 1996.

\bibitem{kosterlitz1974critical}
Kosterlitz, J. M. (1974). The critical properties of the two-dimensional xy model. Journal of Physics C: Solid State Physics, 7(6), 1046.

\bibitem{joyce1967classical}
Joyce, G. S. (1967). Classical heisenberg model. Physical Review, 155(2), 478.

\bibitem{stanley1966possibility}
Stanley, H. E., \& Kaplan, T. A. (1966). Possibility of a phase transition for the two-dimensional Heisenberg model. Physical Review Letters, 17(17), 913.

\bibitem{newell1953theory}
Newell, G. F., \& Montroll, E. W. (1953). On the theory of the Ising model of ferromagnetism. Reviews of Modern Physics, 25(2), 353.
\bibitem{cipra1987introduction}
Cipra, B. A. (1987). An introduction to the Ising model. The American Mathematical Monthly, 94(10), 937-959.
\bibitem{Pathria2022}
Pathria, R., \& Beale, P. D. (2011). 12—Phase transitions: criticality, universality, and scaling. Statistical mechanics, 3rd edition, R. Pathria and PD Beale eds., Academic Press, Boston, USA.
\bibitem{Herbut2007}
Herbut, I. (2007). A modern approach to critical phenomena. Cambridge University Press.
\bibitem{oliaei2023patterns}Oliaei-Moghadam, H., Moodley, C. and Hosseini-Farzad, M. Patterns for all-digital quantum ghost imaging generated by the Ising model. {\em Optics I\& Laser Technology}. \textbf{163} pp. 109392 (2023)

\end{thebibliography}

\end{document}